# A Virtual Processor brings back the Free Lunch*

## Efficient Automatic Parallel Execution of Sequential Tensor Codes: An Engineering Approach

Haymo Kutschbach
ILNumerics GmbH
Berlin, Germany

## ABSTRACT

This work introduces a self-optimizing *virtual processor* (VP) for numerical array programs that shifts parallelization from a manual developer task to a cooperative, agent-like runtime mechanism. Instead of relying on centralized task-graph scheduling, static compiler optimization, or explicitly annotated parallel constructs, the VP uses a decentralized network of cooperative execution segments, derived from the *stream* of numerical instructions and their data dependencies at runtime.

Each segment makes only local decisions about when, where, and how to prepare and execute its computation, including task placement, kernel preparation, and data movement. No central scheduler or mapper instance determines the execution globally; instead, scheduling itself is parallelized and distributed over time – asynchronously and strictly dependency driven.

The overall execution strategy emerges from concurrently executing local segments, continuously responding to data availability, cost estimates, system state, hardware capabilities, and problem size.

While preserving the sequential semantics of the program our VP automatically exploits parallelism across large program regions rather than being limited to individual loop bodies, modules, or explicitly marked parallel sections; developers are not required to design or encode a parallelization strategy.

The current VP primarily targets low-latency strong scaling on local heterogeneous hardware, covering workloads from small, latency-sensitive array operations to large data-parallel computations. The current implementation targets the predefined array instruction set of the ILNumerics.ONAL[22] domain-specific language, while the underlying concept is applicable to general array-based numerical programming models such as MATLAB and NumPy.

* Sutter, H. (2005). The Free Lunch Is Over: A Fundamental Turn Toward Concurrency in Software. Dr. Dobb's Journal, 30(3), 202-210.

## 1 Tackling a Long-Standing Problem

Since the beginning of the “parallel age”, access to increasing computational power is no longer "for free". To increase the efficiency of programs, it is no longer sufficient to simply invest in the next generation of hardware. Today, the increasing performance of hardware goes hand in hand with the increased complexity of its parallel, heterogeneous architecture.

Several tools, technologies, and interfaces today support programmers in the time-consuming *manual* parallelization of an algorithm, which is usually sequential in origin. The success of such endeavors depends on expertise as well as the time and money invested. An increase in performance due to manual adaptation to specific hardware is usually accompanied by reduced compatibility for other hardware configurations. It reduces the maintainability of the software and leads to high costs in many industrial areas. Nevertheless, manual parallelization can only exploit *some* parallel potential from a program. Near optimal solutions can only be realized for specific subproblems of limited complexity. In practice, large parts of the computing capacity of modern hardware remains unused in such applications [3], [4].

*Semi-automated parallelization* ([6], [7], [9], [13]) can be useful if the dominant workload of a program concentrates in a few code blocks. While it leaves the programmer with less manual effort, important decisions about which part of code is executable on which processing unit must still be performed manually at development time. Thus, such attempts derive detrimental aspects of manual parallelization.

*Fully automated* solutions often target tensor or array codes, due to their instruction's limited side effect and aliasing potential. Many array prototyping languages attempt to optimize individual array instruction implementations using SIMD and/or multicore processing on CPUs [15], [16], [17]. Joining subsequent array instructions can help to better address their inherent data parallelism [8], [12], [13], [14]. However, the feasibility of such attempts depends on workload sizes. Since such attempts cannot exploit parallelism beyond individual expressions, they cannot compensate latencies when dynamically switching processing units.

Tools like Legate [1] (based on Legion [2]) and Dask [5] attempt to leverage both on compute clusters: data and task parallelism. But scheduling- and management overhead prevents them from realizing the same advantage on a single compute node. The single-machine case, however, accounts for most applications, many of which need high performance, too.

In the remaining section 1 the introduction of a VP is motivated. The theoretical background and technical details of its inner functioning are provided. In section 2, we compare the efficiency of the presented VP with the traditional execution schemes in NumPy and FORTRAN. Section 3 summarizes the findings and future perspectives.

## 1.1 Introducing the Virtual Processor

When software developers today look back with a bit of nostalgia on the performance increases of the single-core processors of the last 50 years, they remember above all the continuous increase in clock speeds. However, the possibility of accelerating sequential codes through faster hardware has already been realized considerably in the single-core processor through *parallelism*!

The processor provided an interface and guaranteed its validity. (Sequential) programs could use this interface and be confident they would receive a result according to these guarantees. Continuous optimization of the internal functioning of the processor led directly to an increase in performance without any program changes. In addition to increasing the clock speed, the "Free Lunch" consisted of clever methods that optimize the efficiency and interaction of individual components within the processor core. The introduction of cache hierarchies to reduce latency during memory accesses and the parallel execution of instructions by pipelining are particularly well-known representatives.

To enable comparable advantages for modern, heterogeneous architectures, the entirety of a system's devices can be abstracted as a "virtual processor" (VP). It substitutes the conventional CPU for execution of a program on a heterogeneous system. Like the CPU, the VP's interface guarantees correct results for any *sequential* user program. The specific design of the VP's functionality determines the efficiency of the calculation results.

The VP presented in this paper is designed to efficiently execute algorithms of established numerical array languages, such as NumPy, Matlab®, and ILNumerics®. It can be applied to all common numerical languages based on n-dimensional array / tensor objects. A production-ready prototype currently exists as a compiler for the domain-specific, numerical language ILNumerics, embedded in C# /.NET. This language fully supports the semantic and features of NumPy and Matlab®. ILNumerics VP automatically scales the efficiency of sequential array algorithms with varying, parallel hardware capacity.

## 1.2 From Sequential Code to Execution Nets

The overall execution time of a program can be measured as the time between two simple events: 1 - the main thread starts processing, and 2 - the main thread ends processing. Every action performed that helps the main thread finish earlier with the same result correctness can be used as a valuable optimization.

Useful optimizations prevent the main thread from delays and reduce the operations performed by the main thread. The general functioning of the presented VP is designed around these goals. While retaining the *sequential execution semantics* of the program, it attempts to have the main thread finish processing the instruction stream as early as possible. Most time-consuming work required for the correct result is shifted to other computing resources available for concurrent processing.

Instead of calculating the array instruction results sequentially, the main thread integrates each instruction as a new node into a highly dynamic, volatile *execution net* (EN) data structure at runtime. Nodes of the EN enter the structure in the same order defined by the user program, capturing the sequential semantic of the user program. The preliminary initialization of an instruction node and its output array data encompasses the main thread's entire contribution to its processing. Thus, the main thread typically continues integrating a *subsequent* instruction from the user code long before the result of the previous node is calculated.

During the execution of array algorithms, the EN often contains hundreds or thousands of nodes representing *upcoming instructions* and their data, and it links them according to their *dependencies*. Therefore, the EN contains nodes forming individual directed dependency graphs (DDG) corresponding to the user code region already visited by the main thread. For efficiency reasons, nodes often reference more than one array instruction from the user program.

## 1.3 Concurrency by Autonomy

While the main thread steadily adds new nodes to the EN, existing nodes *autonomously* prepare their instructions for processing. This includes deciding which available (sub-)device is best used in which way and when for execution. Nodes start with these scheduling decisions as soon as all required information is available, including input data sizes and available device states. Scheduling also includes providing input data for the instruction(s) of an EN node on the selected device, updating the output data configuration of the node, and informing all nodes receiving the output as dependencies (i.e., parent nodes in the DDG) about state changes.

When all input data arrays for an EN node are configured and available on the selected device, the node triggers the processing of its instruction(s) on this device. It then updates its output with the element values and signals dependent nodes. After a node completes processing, it removes itself from the EN.

The autonomous processing of the nodes leads to *concurrent* processing of many independent nodes from a single DDG or multiple DDGs. The concurrency level depends on the parallel potential in the user code region represented by the current state of the EN. If sufficient independent instructions exist in the array instruction stream, all the algorithms' and the hardware's parallel potential is automatically and efficiently exploited.

After this high-level picture of the VP's general functioning, we will investigate its workings in more detail.

## 1.4 Optimizations in a Virtual Processor

Like a single-core processor, the VP also must optimize the efficiency and interaction of individual components in the heterogeneous system without compromising the correctness of the result. Since no support from a clock speed increase is expected, the VP designer can concentrate on the efficient, simultaneous use of multiple processing units (e.g., multiple cores, GPUs, and the like) to process the nodes of the EN correctly.

Our focus is on the following unguided techniques, which will be addressed:

- Identifying parallel potential (1.4.1)
- Workload scheduling (1.4.2)
- Instruction stream optimizations (1.4.3)
- Array Pipelining (1.4.4)

*1.4.1 Identifying parallel potential,* i.e., code areas that can process data independently, is traditionally performed at development time, utilizing elaborate analyses, often based on directed dependency graphs (DDG). Here, it is desirable to consider as large a code area as possible. In practice, however, the size is quickly limited by increased complexity or by the existence of expressions that can only be evaluated at runtime. Examples are virtual function calls, pointer aliasing and conditional expressions, as the `if`-statement. This is one reason why almost all methods for semi-/automatic parallelization today are limited to certain, often explicitly definable code areas of limited complexity – mostly: simple loop bodies.

The VP presented here takes a different approach. Independencies are identified based on the array instruction stream (AIS), which is processed by the VP at runtime. In the AIS, array instructions of the original user program are lined up in the order of their processing by the main thread and feed the current EN.

If two array instructions can be executed independently, they expose a parallel potential. Recognizing this potential is much easier and more robust at runtime than through demanding dependency analyses at development time. In fact, the dependency analysis in the VP can be reduced to local observations of each individual instruction at runtime. Whether it can be executed independently can be directly decided from the state of the data it receives for processing. This data represents instances of the array variables of the user program. The VP of this work extends common data properties (size, type, element values) by an additional "state" flag. It takes at least two values: *partially configured* and *completely configured*.

The VP defines rules [21] that control the execution of the array instructions based on the state:

1. An instruction whose input data is *fully configured* is independent and can be processed immediately and concurrently with other statements currently executing.
2. An instruction whose input data is only *partially configured* is executed as soon as the data's state changes.

The "dependency" property, thus, translates into a *temporal dimension at runtime*: independent operations are executed immediately. Operations whose data is not yet available will be executed as soon as the input data are ready. This approach supports that i) the original, sequential semantics of the user program are preserved, and ii) that the parallel potential of the algorithm part corresponding to the current EN is fully exploited (subject to sufficient parallel capacity of the hardware configuration).

The nodes in the EN represent array instructions spanning arbitrary user code regions, disregarding function boundaries, code block scopes and conditional expressions. Therefore, the VP identifies and harnesses the parallel potential from *varying levels of code granularity*–a significant enhancement compared to traditional attempts, such as parallelizing loop bodies.

*1.4.2 Workload Scheduling.* The VP gives each node in the EN extensive autonomy to perform important scheduling decisions, especially about which implementation is suitable on which device, and when is a good time to start processing. Again, such decisions are best made at runtime and immediately before starting execution of an array instruction. Currently, not only are all data properties (such as size, locality, and state) available, but all hardware properties are also determined. While many hardware properties can be anticipated at development time or at least at system startup time, other properties cannot. For example, to successfully balance workload throughout multiple devices, their current utilization rate is crucial to consider but obviously not available before runtime.

The autonomy in scheduling decisions directly leads to *concurrent* processing of the nodes in the EN [21].

*1.4.2.1 Latency Optimization.* Another important aspect of workload scheduling in a heterogeneous system is minimizing the detrimental effects caused by latencies. Here, "latency" refers to the delay caused by switching processing units. It typically includes a delay for data transfer to/from dedicated memory (if required) and a certain overhead of starting processing on another unit.

One preferable method of dealing with latencies is to *hide* them as much as possible. The presented work hides latencies for the main thread by asynchronous instruction processing. Instead of waiting for the completion of a result, the main thread continues visiting subsequent array instructions. Thus, it continues doing 'useful' work without stalling.

On a lower level of the processing of EN nodes latencies are introduced by determining and preparing the best execution strategy and by starting processing on the most suitable device, including potential memory transfers. Such actions are unavoidable for a virtual processor. They incur certain runtime overhead, regardless of the efficiency of the concrete implementation. Therefore, the presented VP considers associated costs when making scheduling decisions.

For each device, the VP maintains a measure of the overall cost of processing all instructions scheduled, similarly to a device specific processing queue. This "cost ahead" is typically idealized and based on the minimal processor cycles required to complete the array instruction with specific input data. Additionally, normalizing this value with the actual clock frequency of the individual device makes costs comparable among multiple devices. The "cost ahead" measure is used to trade expected latencies for crossing device boundaries against the expected time required to "wait in the line" of already enqueued instructions on the current device [19].

These specific scheduling decisions are performed as soon as the problem size of an instruction becomes available (see below: Array Pipelining). Typically, this time comes significantly earlier than the start of calculating the result's element values. It allows the VP to trigger required memory copies ("cross device prefetching") and/or kernel code compilation as early as possible. The actual copy operation is performed asynchronously from the current node's scheduling thread, which helps hiding associated latencies.

Further, scheduling is performed concurrently to the main thread by individual nodes of the EN and on individual worker threads, which further reduces the impact on the whole system (because actions performed by other threads at least partially hide latencies on one thread).

In summary, all these precautions allow the VP to almost hide the latency of crossing device boundaries completely. This ability marks another significant advantage over traditional methods of addressing heterogeneous device configurations.

*1.4.2.2 DDG Scheduling.* In the presented work, an independent instruction node autonomously schedules its workload for processing on the most suitable device (see above). Such instructions can be seen as the leaves of the corresponding directed dependency graph (DDG), including the instruction. *Dependent* instructions, conversely, correspond to the DDG's inner nodes.

An EN potentially contains many DDGs, corresponding to independent code regions. For example, each iteration of an iteration-independent loop body in user code leads to an individual instance of a corresponding DDG in the EN. As we have seen earlier, nodes corresponding to independent DDGs are executed concurrently. Moreover, each independent node's execution is performed within its own *temporal context*: when the node is to start processing is determined by the at least one node processing its dependency. Therefore, the method scales well with multiple or many concurrent branch executions and effectively minimizes unwanted *contention issues* during processing on many processing units.

Similarly, leaves within the same DDG may also execute concurrently. Whether or not concurrent processing of such leaves is advantageous depends on many influencing factors, including the shape of the input data processed. For details refer to [20].

EN nodes in the ILNumerics VP delegate scheduling of their workload as 'continuation' to the node that completed processing of the last actual input data dependency. Therefore, subsequent nodes within a DDG branch share a common computing (sub-/) device unless cost-based scheduling instructs otherwise. This rule reduces processing overhead and contention by preventing frequent switching of processing contexts within a DDG branch. It is important for the efficient processing of small and mid-sized data.

*1.4.3 Instruction Stream Optimizations: micro-JITs.* We have seen how the nodes of the EN are arranged, and how their workload is distributed to individual, parallel computing devices. However, the processing on the selected device requires optimization, too. The optimal implementation of an array instruction *varies* with the specific device and with data properties. Therefore, suitable nodes in the EN carry an intermediate code which is just-in-time (JIT) transformed and JIT-optimized for the selected device and for the specific data properties observed.

This finalization is performed by so-called 'micro-JITs'. The 'micro' in their name indicates one crucial goal of such JIT compilers: to minimize runtime execution overhead. Therefore, each micro-JIT specializes in exactly one instruction tree, corresponding to the specific user instructions in each node of the EN.

Each micro-JIT encompasses a *parameterized expression tree*. When executed, it forms an optimized kernel implementation of the user instruction(s), taking into consideration all relevant hardware and data properties.

Micro-JITs typically apply all common and essential optimizations when generating kernels for a node: loop unrolling, tiling, cache-awareness, and SIMD vector utilization. Further, the optimizations applied by a micro-JIT and the low-level instructions used are automatically adapted to the target device type (CLR/ CPU/ GPU/…).

*1.4.3.1 Kernel Staging.* High specialization for kernels at runtime promises highly efficient code. On the other hand, it requires adopting the final code whenever at least one relevant data property changes. In some situations, delays due to frequent re-JITing may be undesirable. Hence, the micro-JITs automatically adopt the level of specialization of kernels. The presented VP starts with a general base kernel implementation with minimal optimizations. When processing statistics indicate feasibility, this version is more and more optimized and specialized in subsequent runs. When a node attempts to start processing a kernel of a certain specialization stage that is not (yet) compiled, the node's micro-JIT begins preparing the required kernel asynchronously. Meanwhile, the node continues processing using a kernel from the next lower specialization level until the more optimized version is ready.

*1.4.3.2 Memory Management.* When optimizing memory accesses, the goal is to prevent excessive memory utilization. The best way to avoid memory access delays is not to access the memory. This simple truth holds for a virtual processor in the same way as for traditional methods. The 'how' is where things become intriguing.

It would be clear for a compiler expert that implementing a numerical language brings its own requirements regarding efficient memory management. Methods as to use views for referencing the same memory storage instances for individual arrays, lazy copy on write, and arbitrary strided storage schemes are helpful for any numerical language implementation. In the case of a domain specific language (DSL) in a strongly typed, garbage collector (GC) managed host environment such as .NET, i) an automatic deterministic disposal of array instances, ii) distinguishing array instances by types representing their lifetime, and iii) reusing ample memory storage via a memory pool, add to this list.

The following will focus on memory optimizations, specific to the presented virtual processor.

*1.4.3.3 Removal of intermediate results* is realized by joining (or: fusion or merging) of adjacent nodes within the same DDG branch in the EN, if appropriate. Thus, intermediate results are removed, limiting the overhead of memory access. Interestingly, the resulting efficiency is subject to the shape and size of the input data arrays and the instruction [20]. For example, a binary array expression performing broadcasting may complete faster using an intermediate array as input instead of redundantly calculating many identical input element values along a broadcasted dimension.

Thus, the VP may join the same instruction with its neighboring instruction in one processing run. But it may not be joined in a following invocation with changed data properties. Consequently, the segmentation of the user code may *change* frequently during execution at runtime and so does its low-level kernel implementation.

*1.4.3.4 Multi-Location Storage.* A user of the numerical language of the presented VP uses specialized array classes to represent numerical arrays / tensors. Instances of an array not only store the size, shape, storage layout, element type, and state of the tensor. The VP further transparently adds a location property and several utilization counters. Arrays allow to store and maintain their element values on any memory accessible by a processing unit. At runtime, the VP manages the memory transparently to the user: memory of element data is allocated, copied to/from any device memory, and invalidated on write access during scheduling of nodes on arbitrary devices.

*1.4.3.5 Memory Coherence*. As explained earlier, nodes of the EN start processing when all input data are 'ready'. This respects any processing dependency on the data's *write* state (read-after-write and write-after-write dependencies).

However, during concurrent execution, multiple nodes from the EN may access the same ready array instance concurrently, for reading (input and input/output data) and writing (input/output and output data) as defined by the user code. Thus, the VP must also guarantee correctness for concurrently accessing data defined in read-after-read and write-after-read (WAR) order. Here, the only critical access pattern is when a node starts writing to an array's storage used for reading in a former processing which is not yet completed (WAR).

The VP of this work handles WAR conflicts by *array renaming.* When a node starts preparing its array parameters for processing, it uses a *copy* of such parameters that are still known to be used in a pending read operation. This method exposes some similarities with register renaming for Out-Of-Order execution in a CPU. However, its implementation is more challenging because we deal with n-dimensional and volatile data. The renaming involves a copy operation, potentially accessing multiple device memories, and in general, must be implemented thread-safely.

Array renaming supports minimizing contention during concurrent processing of the nodes in the EN. The node attempting to write to an array must not wait for all read accesses to complete. On the other hand, it may increase an algorithm's memory footprint, emphasizing the need for an efficient memory management. Note that any latencies introduced by the additional copy operation are often hidden by useful work concurrently performed by other nodes.

*1.4.4 Array Pipelining.* We have shown that EN nodes` autonomous processing can leverage parallel potential among subsequent array instructions. But there is also parallel potential *within* array instructions! Using it for parallel processing requires breaking down the closed system of an array instruction and considering its individual components. Therefore, nodes in the EN are signaled by nodes, processing their dependencies *several times*, each corresponding to a *partial completion* stage [21].

The presented VP splits the processing of array results into three stages: 'size completed', 'device configured', and 'result completed'. When a node (A) starts processing it first configures the size of its output array instance (C). For many array instructions, the output size only depends on the size of its inputs. Thus, the output size calculation can be done early in the nodes processing chain.

Subsequent nodes having the output (C) as a dependency start processing *their* result (size) based on the partially configured input (C) and afterwards inform their dependencies, too. Similarly, nodes inform subsequent nodes about completing scheduling decisions and partial configurations of output array instances (i.e., size and memory locality) by signaling the state: 'device configured'.

Thus, array data referenced by nodes in the current EN are configured in multiple stages, and each partial configuration is populated through the EN early on. As a result, nodes start processing earlier, kernels become configured earlier, memory copies start earlier, kernel compilation starts and completes earlier, and kernels execution starts and completes earlier – subject to sufficient parallel capacity of the hardware.

## 1.5 Putting it All Together

In the following, the execution strategy of the presented VP is demonstrated by means of a simple, concrete code example, which might be part of a larger program:

```
R1 = abs(fft(data1));
R1 = sum(R1 + abs(fft(data2)), 1);
```

Here, data1 and data2 are predefined matrices, and R1 is an array / a tensor variable holding intermediate results and the result of this short algorithm. The algorithm reduces the columns of the result by adding the magnitude values of the Fourier transform of both input data. Its (arbitrary) purpose is to demonstrate the execution of common array instructions. Its mathematical meaning is not relevant here.

The algorithm exposes some parallel potential for concurrent execution, as seen from the directed dependency graph (DDG) in Figure 1.

We will now 'execute' this algorithm on an imaginary system comprising a multicore processor with four logical processor cores. In the following, we focus on automatically distributing workload tasks onto multiple processing units. For simplification optimizations of the instruction stream (segmentation and array pipelining) are not demonstrated.

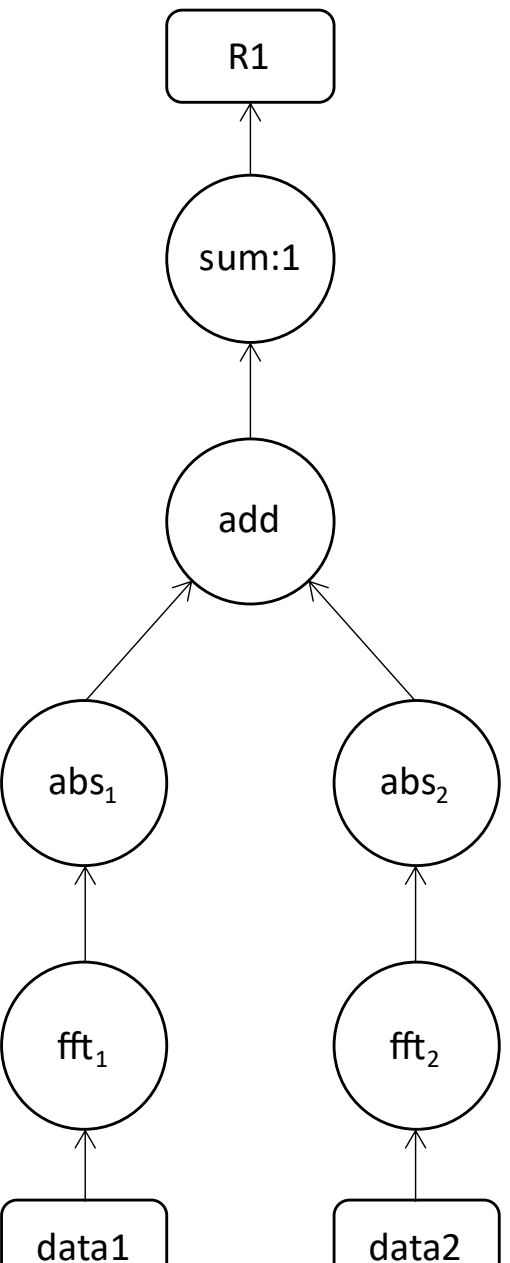

**Figure 1: DDG of the example algorithm. Rectangles correspond to data variables; inner nodes represent instructions. Edges mark dependencies.**

In a traditional approach each instruction is processed 'synchronously', i.e., the main thread *starts and completes* the processing of each instruction. It starts processing subsequent instructions only after the current instructions result is completed. The completion of the whole algorithm is marked by "End".

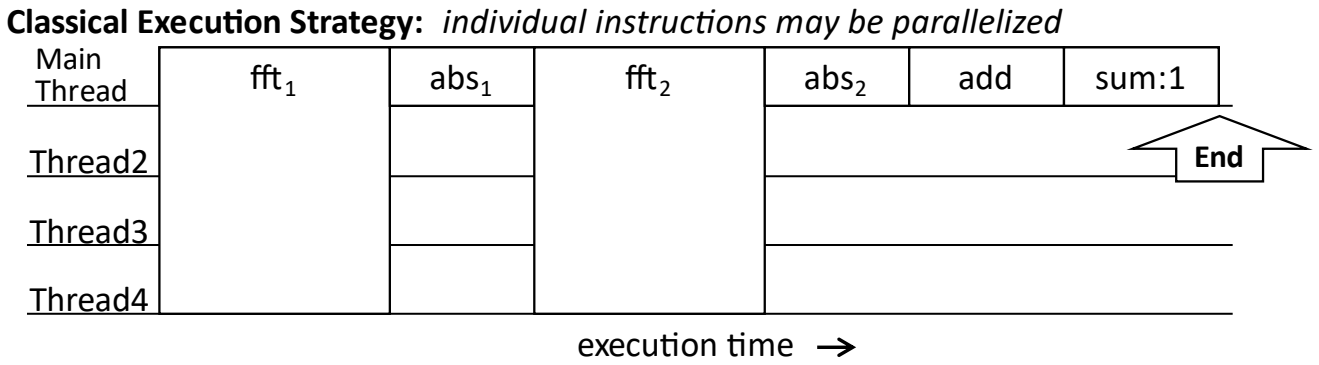

**Figure 2: Traditional execution strategy. Array instructions are processed in the order defined by user code. Some instructions may use multiple processing units.**

Often, the main thread involves further processing units to calculate the result(s) of individual instructions. For example, the FFT instruction in the execution runtime of most numerical languages uses third-party libraries, as Intel's Math Kernel Library (MKL), for processing. The MKL typically attempts to parallelize heavy workloads by using multiple threads. Therefore, the $fft_1$ and $fft_2$ instructions in Figure 2 use all four available threads for processing.

While this approach can speed-up large workloads, it involves crossing device boundaries (threads / cores) on instruction scope. Latencies must be compensated by the additional processing capacity of the processing units. This, in turn, requires the workload of *each parallel chunk* to exceed certain thresholds, which is often hard to anticipate in advance.

Further, for completing an instruction, the main thread must commonly wait for completing processing all chunks. Thus, additional latencies due to synchronization are introduced. Moreover, most simple instructions use only one single thread, leaving precious computing resources idle.

A potential situation at runtime when using our VP is presented in Figure 3. The *main thread* visits each instruction in user code in the regular order defined in the program. It adds each array instruction as a new root to the corresponding DDG in the EN, according to its input arrays. Further, any output array of the instruction is initialized – without configuring element values or size yet.

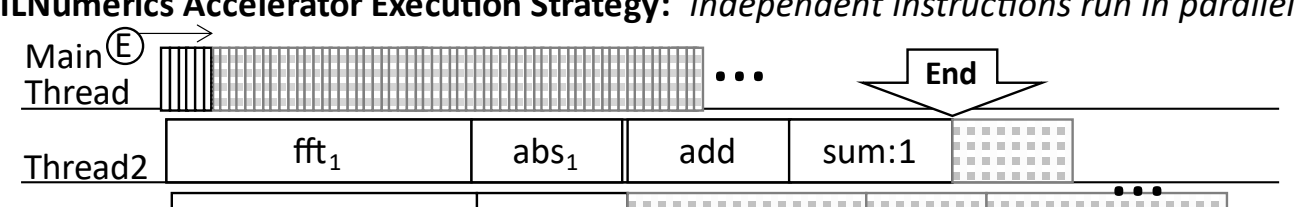

**Figure 3: Thread view in a virtual processor. Many array instructions from the EN start processing concurrently. Processing units are used as available.**

After this comparably quick operation the main thread immediately proceeds with subsequent instructions. Such (arbitrary) instructions are defined in user code *after* our algorithm and are marked in Figure 3 by a gray pattern.

The circled E in Figure 3 marks the point in processing time when the main thread has visited all instructions in our algorithm. However, while the main thread initialized a result array R1 from the last instruction, neither R1's element values nor size have been configured yet.

While the main thread performs the above actions, the nodes in the EN start processing *autonomously*, subject to the state of the input data for a node. For this example, we assume that data1 and data2 are already fully configured. Further, we assume, that the node corresponding to instruction $\text{fft}_1$ starts first. Shortly after, the node corresponding to instruction $\text{fft}_2$ begins processing. A node automatically considers the state of all processing units and enqueues itself to the command queue of the most suitable processing unit. Hence, both instruction nodes: $\text{fft}_1$ and $\text{fft}_2$ are processed by individual processor cores in parallel: thread2 and thread3 in Figure 3.

Note that, in contrast to traditional execution schemes, the processing of the fft instruction is often performed *single-threaded*, removing unwanted contention and other detrimental effects for small- and mid-sized data.

When $\text{fft}_1$ completes processing on thread2 the completion of its output array signals subsequent nodes from the EN's DDG: $\text{abs}_1$. The $\text{abs}_1$ instruction continues processing on the same thread to prevent from unwanted latencies when crossing a device / thread boundary. Similar actions are performed by thread3 for the $\text{fft}_2$ and the $\text{abs}_2$ instructions.

When both $\text{abs}_1$ and $\text{abs}_2$ instructions have been completed, the node corresponding to the add instruction is signaled for processing. Its completion triggers the processing of the reduction sum along axis1 on the result of add. The sum instruction completes the configuration of the R1 result array, which had been initialized by the main thread earlier.

With the completion of the sum instruction node on thread2 our example algorithm part is finished, marked by the fat "End" arrow in Figure 3. This end is located earlier than in the traditional approach. However, it is far behind the main thread, which – at this time – progressed forward by many (potentially thousands of) further array instructions, following our example partial algorithm in a real-world program. All these instructions have also already been integrated into the EN as DDGs. Some leaves of these EN DDGs had already started processing – concurrently to the nodes from our algorithm. They have been filling formerly idle slots of our CPU cores.

The CPU, hence, sees a good utilization as is marked in the schematic view of Figure 3 by grayed rectangles in all worker threads. What is demonstrated here with a work stealing thread pool on a CPU, works similarly on other devices using command queues. The VP gains speed-up by using multiple processing units to process subsequent, independent instructions concurrently.

## 2 Advantages in Practice

We present studies performed to demonstrate and quantify the efficiency of the presented VP. The selection of experiments focuses on measuring the impact of new optimizations enabled by a VP and comparing the results with the efficiency of running the same code without these optimizations, using a traditional approach on the same platform.

We begin our investigation with a preliminary experiment designed to establish a rough baseline and assess our current position in the field of other popular computing languages, their compiler optimizations, and their runtime technologies.

The artifacts accompanying this paper are archived at Zenodo: https://doi.org/10.5281/zenodo.20407801

The development repository is available at: https://github.com/hokb/decentralized-array-execution-artifacts2026

### 2.1 Speed Compared to NumPy & FORTRAN

This study evaluates the presented VP's efficiency compared to the popular NumPy language, specifically analyzing how its fully automated optimization performs against similarly user-friendly technologies. The results are benchmarked against a FORTRAN implementation as the performance baseline.

**Expression to measure:**

```
sum((m0 & (A << shift)) | (~m0 & B), dim: 1)
```

**Data used:** `A` and `B` are arrays with 32-bit unsigned integer elements and four dimensions of lengths: [507,10,5,17] and [1,1,5,17] respectively. `m0` and `shift` are scalar constants.

**Measurement method:** for each technology (NumPy/Python, ILNumerics, FORTRAN), the expression is evaluated repeatedly in a non-optimized program and a program using full optimization. Each sample represents the average execution time consumed by 1000 invocations throughout the first 10 seconds of running the application. Warm-up phases are included in the measurements, to learn about the impact of JIT compilation. Details of each measurement are provided in Part 1 of the artefact description appendix at the end of this document.

**Measurement system** (used for all subsequent experiments):
RAM: 64 GB DDR4, CPU: i7-12700K, Win 11 Pro, .NET 8.0

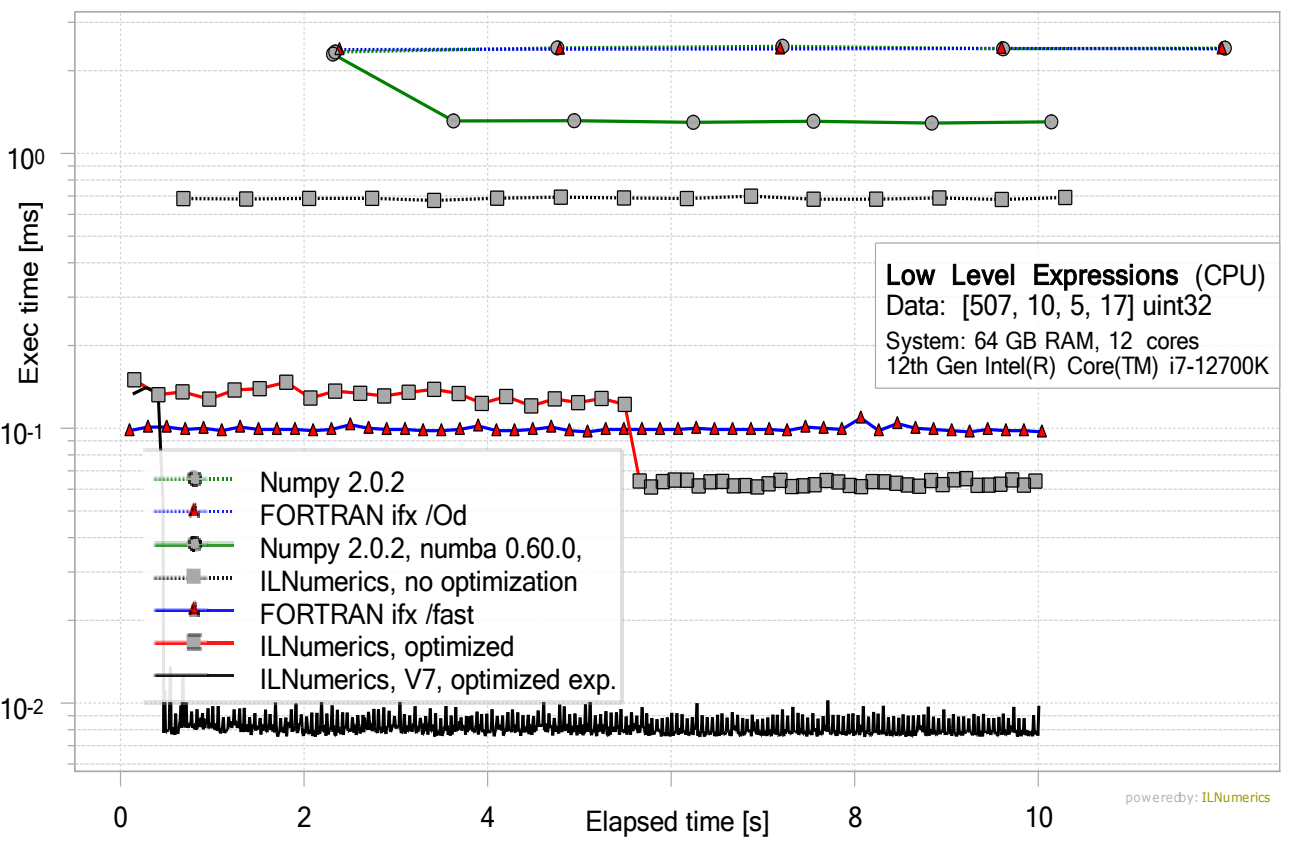

**Figure 4: Execution speed of a non-toy array expression in ILNumerics, NumPy, and FORTRAN over the app's run time. Unlike Numba and FORTRAN, the VP accelerates many instructions simultaneously across multiple units.**

**Discussion:** The execution speed measurements are shown in Figure 4. Interestingly, FORTRAN without using any optimizations (ifx /Od flag) was slowest (~2.3 ms). It is almost on-par with NumPy (~2.2 ms). The Numba JIT compiler improved the NumPy speed by nearly factor 2 (1.2 ms).

The non-accelerated version of ILNumerics (using only the .NET JIT) required ~0.7 ms for each expression evaluation. The presented VP methods accelerate the same code by a magnitude (~0.06 ms). It is now reliably faster than the Intel® FORTRANs ifx compiler, having all optimizations enabled (/fast flag, ~0.1 ms).

All measurements were performed using a multicore CPU only. No other processing units (GPU) were involved. Enabling a specializing feature in the presented VP (experimental in version 7.1.22) makes the potential for future optimizations clear, even when considering only CPU optimizations. On the CPU alone the same code now runs faster than FORTRAN by a factor >10.

Note further, how the impact of JIT compilation is visible from the plots. Both NumPy using Numba and ILNumerics with VP acceleration required around 3 seconds to compile an optimized runtime kernel for the first time. However, the JIT compiler in ILNumerics Accelerator takes advantage of kernel staging and uses dynamic, precompiled, parameterized expressions (see: 1.4.3.1) to reduce JIT compilation latencies. Hence, the observed delay is only effective for the first expression compiled and for loading all related technologies into RAM. Subsequent expressions compile much faster, as seen in the "ILNumerics,V7,optimized exp." experiment. Here, JIT compilation finished in less than 100 ms.

## 2.2 Enabling Robust Loop Parallelization

Many traditional parallelization approaches today center around improving program performance by executing *loop iterations* concurrently. They require at least partial *iteration independence* of the code defined as the loop body and often introduce unwanted latencies due to necessary task delegation and synchronization. Further, all objects accessed by individual iterations must be secured for concurrent access from multiple processing units.

In difference to that the presented VP *retains the sequential execution order* of *dependent* array instructions – also within loops. It renders elaborate independence analyses obsolete. This study investigates its efficiency compared to traditional, manual loop parallelization. To make things more interesting, we start with an *embarrassingly simple* parallelizable loop example to compare the VP's speed with a common and efficient 'Parallel.For' optimization. Afterwards, we investigate the behavior of both techniques for a loop body exposing iteration dependence.

Consider the following code of a simple loop, creating a vector B based on a predefined matrix A in ILNumerics:

```
for (int i = 0; i < A.S[1]; i++) {
  B[i] = sum(abs(sin(A[full,i])), dim:0);
}
```

Note that the right side of the array expression forming the loop body is arbitrary. Here, it was selected to represent a non-trivial workload without introducing (costly) high-level functions, which are also often imported from native libraries.

For years, such loops were frequently used to accelerate high level expressions on multi-dimensional arrays. A programmer might use the above loop to speed up the execution of the formula:

```
B = sum(abs(sin(A)), dim: 0);
```

By manually controlling the iteration of the data, advantageous iteration orders can be enforced, and/or parallelization can be applied. Here, one could use multiple cores and have each core calculating an individual range of elements of B in parallel. In pre-VP times, such parallelization was often achieved explicitly by enhancing the loop optimization with a `Parallel.For` expression from the 'Task Parallel Library' (TPL):

```
Parallel.For(0, A.S[1], i => {
  B[i]=sum(abs(sin(A[full,i])),dim:0);
});
```

In an experiment (Part 2a), we have compared the execution times of the original high-level array expression and its manual loop iteration (each with and without VP acceleration) with the `Parallel.For` version of the manual loop.

**Discussion:** The VP accelerated version of the high-level array expression (without loop) could best use the parallel capacity of the multiple cores (Figure 5). It even beats the manually optimized version using loop iteration and `Parallel.For` parallelization – without any guidance. Note, that both: the 'vectorized' high-level expression and its sequential loop counterpart are roughly on par *without* VP acceleration. However, the VP compiler could speed up the high-level expression on the whole data much better than the manually iterated loop. This is because the manual loop (prematurely) anticipates many decisions regarding the execution strategy – as the parallel grain size. Such decisions are more efficiently made at runtime by considering the entire system.

The learnings from this part: A VP encourages the use of high-level codes that are agnostic to the execution hardware. This recommendation fits seamlessly into the high demands for efficient development of even complex codes while maintaining their maintainability.

Appendix Part 2 contains another code example showing how the presented VP runs high-level FFT codes almost three times faster than the internal multithreaded parallelization within Intel® MKL on the same system.

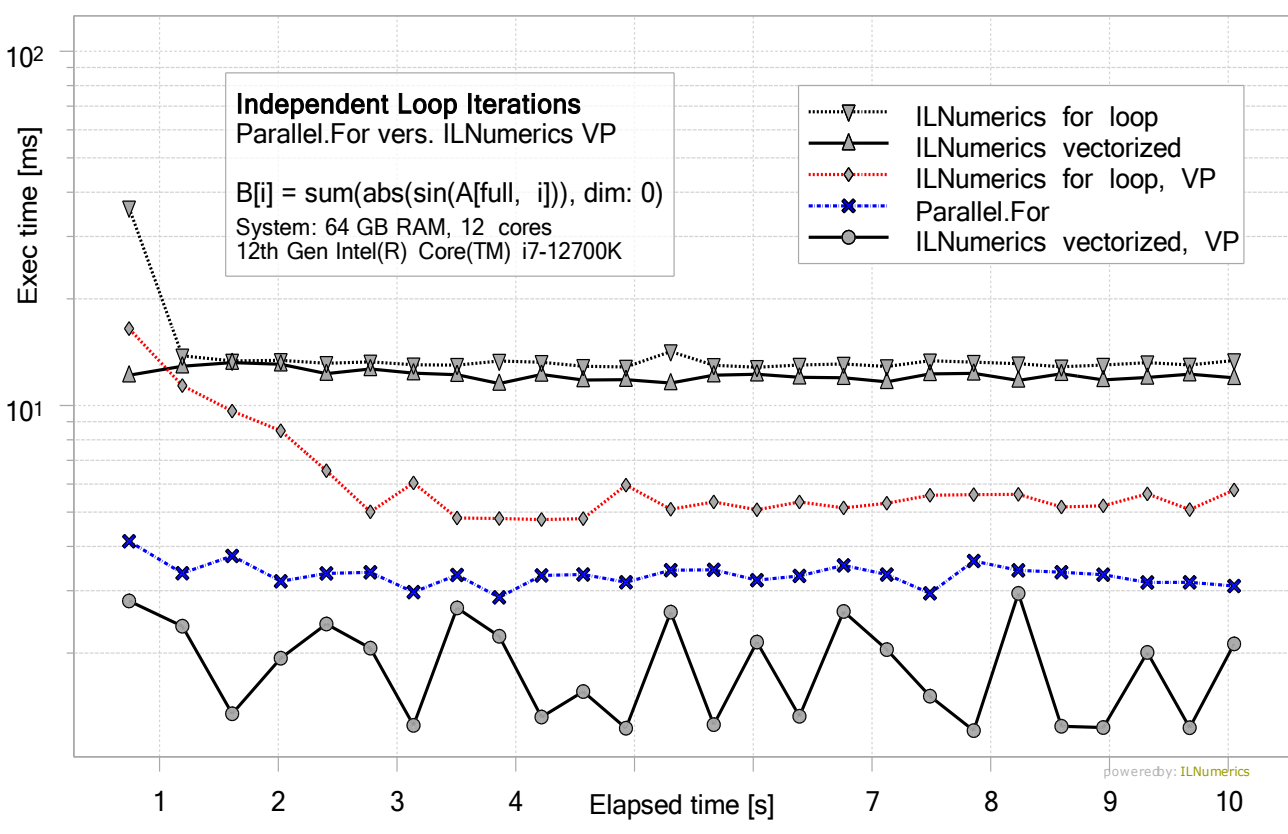

**Figure 5: Various attempts to accelerate a high-level array expression. Manually iterating over a problem dimension ('for loop') can have negative effects on execution speed. Fastest result: the unsupervised VP acceleration of the full high-level expression without outer loop.**

Part 2b investigates the consequences of operating loop iterations that are *not* embarrassingly simple to parallelize. Therefore, we modify the former loop expression to include an iteration dependence. The result now depends on the correct order of write operations to the output array:

```
for (int i = 0; i < A.S[1]; i++) {
  B[r(0,i)]=sum(abs(sin(A[full,i])),dim:0);
}
```

We repeat the measurements from Part 2a with the modified loop. Since we are unaware of an equivalent vectorized version, we focus on measuring the given loop body – with and without VP acceleration. Additionally, we measure the '`Parallel.For`' version. While it cannot produce correct results, it provides a rough guess of a potential speed-up.

**Discussion:** From Figure 6, the VP was able to accelerate the iteration-dependent loop almost to the same degree as an iteration *in*dependent parallelization could perform. This is possible because any delaying effects due to data dependencies are limited to the instructions introducing the only dependency. The VP automatically respects any data dependency (here: the order of writes to the output array in each loop iteration). However, all other (independent) operations are still parallelized on a fine level of granularity and execute concurrently. Thus, large parts of the loop iterations run in parallel. They compute the results of the right side in the array expression in parallel. These results are readily provided to the assignment instruction and only *assigned* to the output array in *sequential order*.

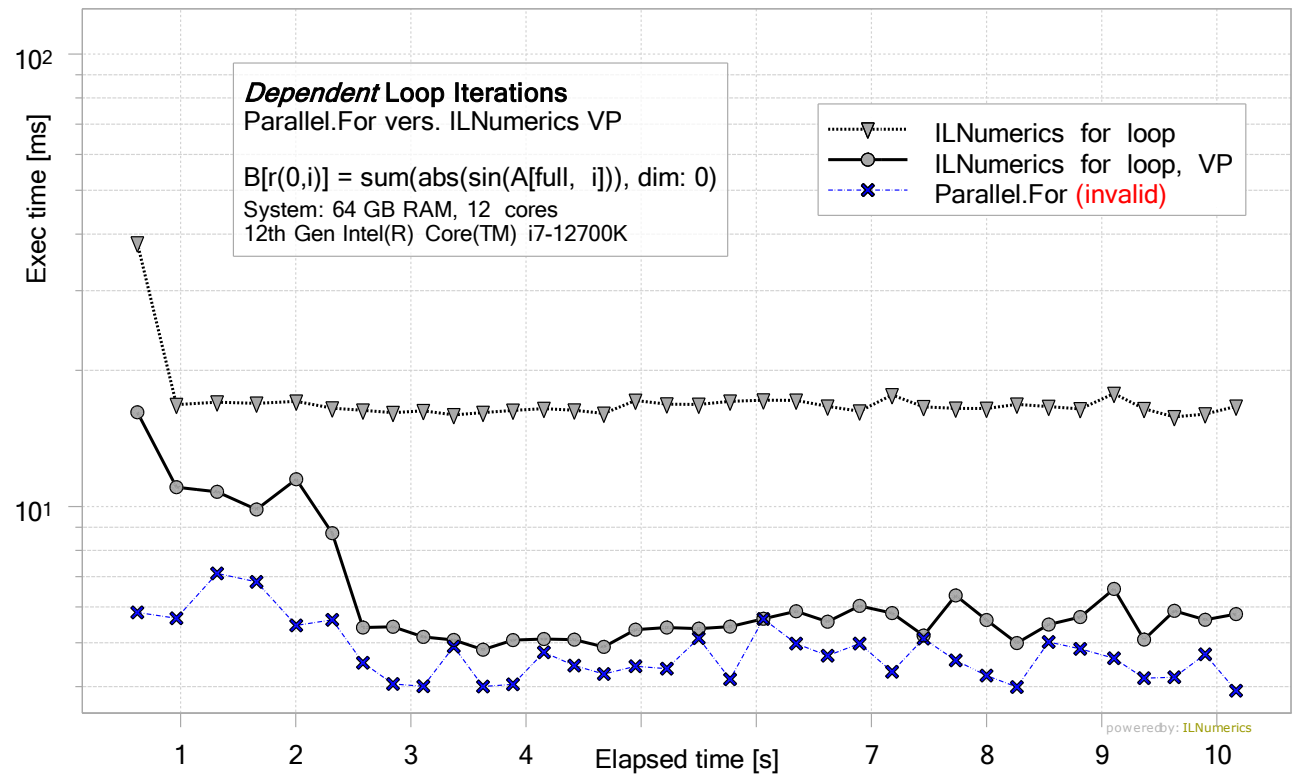

**Figure 6: Accelerating a loop with iteration dependence. A VP respects data dependencies – inside *and* outside of loops. Still, the loop body is accelerated to almost the same degree as if no iteration dependence existed.**

Takeaway from Part 2b: The presented VP can leverage parallel potential even from algorithmic parts that cannot be easily addressed by traditional parallelization methods.

## 2.3 Whole Program Optimization: K-Means

The ability to leverage parallel potential on various levels of granularity of code and hardware components promises to reveal and use a lot of optimization potential that formerly could not be used efficiently. While we traditionally had to focus on those (few) bottlenecks that could feasibly be improved through manual optimization, a VP automatically handles all array instructions in a program – even outside of loops and across function boundaries.

In Part 3, we investigate the VP's performance advantage to a larger algorithm with practical importance: K-Means [18]. This popular clustering algorithm comprises of two main loops: the first loop assigns provided data points to clusters based on a distance measure. The second loop recalculates the cluster centers based on its data members. This *iterative* algorithm is repeated until a stable assignment has been found.

Appendix Part 3 contains the codes of the K-Means algorithm experiment. It compares equivalent versions written in ILNumerics, NumPy, and FORTRAN, respectively. The non-optimized code was compiled using the standard compilation on each platform and compared with a compilation, leveraging automatic optimization – as far as admissible. For NumPy, however, no automatic optimization was applied; Numba and JAX were found to require significant modifications to the algorithm, hence missed the study's 'automatic' requirement.

The FORTRAN code had to be adjusted to exactly match the workload and to produce identical results to the array codes on NumPy and ILNumerics. While the resulting code is ~ twice as long and less readable, it was included in this experiment for benchmarking reasons.

K-Means was run for each technique with the same input data, performing exactly 5 iterations per invocation. The invocations were repeated for 15 seconds. The execution time of each invocation / solution was measured. Results were checked for identity. Times are displayed in Figure 7.

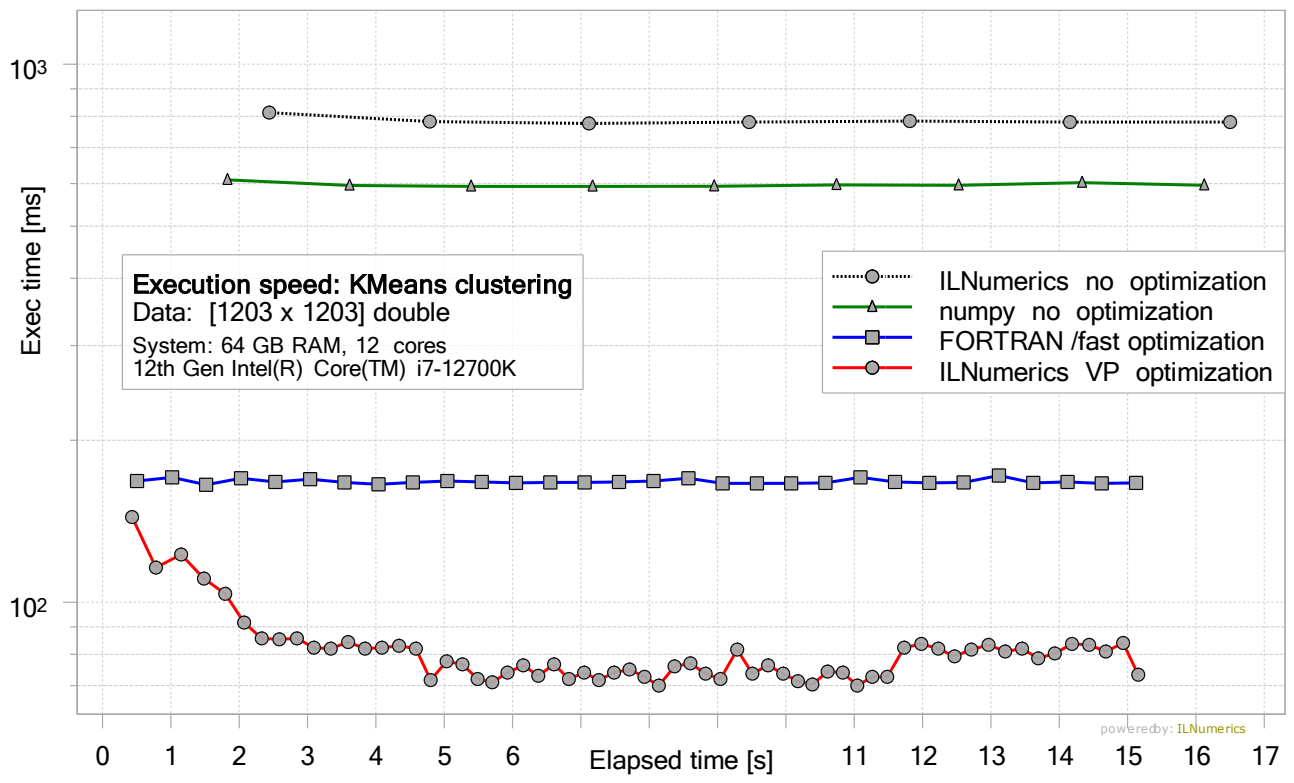

**Figure 7: Execution times of K-Means clustering over app's running time in ILNumerics, NumPy, and FORTRAN. The VP was able to optimize the unmodified high-level algorithm, even beating FORTRAN by a factor > 2**

**Discussion:** Figure 7 shows that the unguided optimizations in the VP accelerate the execution of the K-Means algorithm at moderate problem sizes by roughly a magnitude. The speed advantage is attributed to low-level micro-JIT- and to array pipelining optimizations. Despite the iterative nature of the algorithm, most parts of individual iterations of the dominant loop 1) were executed concurrently on multiple CPU cores.

## 3 Challenges, Limitations, Outlook

A VP brings enormous potential for highly efficient and fully automated use of modern computing hardware, which is inherently parallel and heterogeneous.

The presented work is capable of efficiently processing array-based languages. What may look like a limitation at first becomes less pronounced when considering that the efficiency topic is most compelling for long-running programs, which often handle big data or complex tasks. Numerical array/tensor codes play an essential role in this field. They allow the author to create high complexity without great effort. The availability of low-cost access to high performance might even increase their popularity.

However, a VP is a complex piece of software. At the core, it relies on volatile data structures, being processed and modified concurrently. In theory, the efficiency of this processing is only limited by the capabilities of the available hardware. Such resources, led by memory, are limited. Thus, the designer of a VP is responsible for managing limitations in scalable, reliable, and efficient ways. Once the higher resource demand imposed by highly concurrent processing can be satisfied, a VP uses parallel hardware more efficiently than manual plans.

### 3.1 Summary

We presented a virtual processor capable of efficiently executing general sequential array codes using common, heterogeneous parallel processing hardware. Numeric parts of the user program are transformed into the volatile, multi-temporal space of an EN, formed by a multitude of DDG-associated execution branches.

Each independent node of the EN – autonomously and thus concurrently to other nodes – triggers the processing of an individual DDG branch on a suitable processing unit. The complex interplay of processors, memory, and algorithms is constantly optimized by the VP on various levels of hardware and code granularity.

The presented VP executes sequential array algorithms with superior speed and hardware agnosticism compared to legacy approaches. It often outperforms manually optimized solutions while eliminating the need for developer guidance. The introduced VP could prove to be a breakthrough for the automated, efficient utilization of modern heterogeneous parallel hardware architectures, addressing long-standing challenges in leveraging multicore CPUs, GPUs, and other accelerators, especially in technical fields.

## Appendix: Artifact Description

The artifacts accompanying this paper are archived at Zenodo: https://doi.org/10.5281/zenodo.20407801

The development repository is available at: https://github.com/hokb/decentralized-array-execution-artifacts2026

## Artifact Description (AD)

### 1 Overview of Contributions and Artifacts

*1.1 Paper's Main Contributions*

$C1$ A unguided, self-optimizing Virtual Processor (VP) for array/tensor languages that dynamically distributes workloads across heterogeneous hardware at runtime.

$C2$ The volatile Execution Net (EN) data structure capturing array instructions and dependencies to expose fine-grained parallelism.

$C3$ Autonomous, per-node decisions for task placement, scheduling, latency hiding, and load balancing across multiple processing units.

$C4$ Adaptive, low-level, hardware agnostic Micro-JIT kernels with loop fusion, pipelining, and staged SIMD/ vector specialization optimizations.

$C5$ Automatic array pipelining, fusion, and array renaming to enhance memory efficiency and to ensure correctness guarantees.

$C6$ A prototype implementation in C#/.NET demonstrating automatic, significant speedups over NumPy, FORTRAN, and legacy ILNumerics® on multicore CPUs.

*1.2 Computational Artifacts*

$A1$: Benchmark: Low Level Array Expression
sum((m0&(...)),dim:1)

$A2$: Benchmark: “Embarrassingly Parallel” Loop
for i in ... B[i] = sum(abs ...

$A3$: Benchmark: Loop with Dependencies
for i in ... B[r(0,i)] = sum(abs ...

$A4$: Benchmark: Full Program, "K-Means" Full algorithm (cluster assignment + centroid update)

DOI (all artifacts):
https://doi.org/10.5281/zenodo.20407801

*1.3 Artifact ID Contributions*

| Artifact ID | Contribution | Paper Element |
|---|---|---|
| A1 | $C1$, $C3$, $C4$, $C5$, $C6$ | Figure 4 |
| A2 | $C2$, $C3$, $C5$ | Figure 5 |
| A3 | $C2$, $C3$, $C5$ | Figure 6 |
| A4 | $C1$,$C2$,$C3$,$C4$,$C5$,$C6$ | Figure 7 |

### 2 Artifact Identification

*2.1 Computational Artifact **A1***

The artifact benchmarks the integer array expression:
sum((m0 & (A « shift)) | ( m0 & B), dim: 1)
The efficiency of the computation is compared to equivalent implementations of NumPy (w/ & w/o Numba), and FORTRAN.

#### Relation To Contributions

Artifact $A1$ demonstrates VP’ s efficiency (C1) via autonomous scheduling (C3), micro-JIT optimizations (C4), and fusion/ pipelining (C5), achieving further speedup by crossing CPU core boundaries (C6).

#### Expected Results

The presented VP is expected to pick-up any array / tensor code from the supported language and to execute the instructions more efficiently than FORTRAN and NumPy. Highest execution speed requires efficient lowering of the array expressions for the selected CPU, efficient utilization of more processing capacity and an efficient interplay of all hardware components. Even with a moderately optimized low-level kernel the VP's execution speed (‘ILNumerics optimized’ in Figure 4) is expected to surpass the speed of the other technologies. Applying all optimization features (‘ILNumerics V7, optimized exp’ in Figure 4) should lead to highest execution speed in the benchmark.

#### Expected Reproduction Time (in Minutes)

Artifact setup: 10 min, execution: 2 min, analysis: 1 min.

#### Artifact Setup (incl. Inputs)

Please consult Readme.md contained in the artefact package. An up to-date version is found at:
https://github.com/hokb/decentralized-array-execution-artifacts2026

#### Hardware (all artifacts).

The VP's ability to autonomously involve multiple processing units for computation is demonstrated in the provided VP for multiple CPU cores only. Hence, the artifact can be reproduced on any single-node setup, equipped with at least one multicore CPU. No other processing unit is considered by this artifact.
Software (all artifacts). A Debian based docker image is provided, containing all artifacts. Preinstalled software: Debian 11, gfortran 10.2.1, .NET 6.0. The evaluation is performed inside the docker container. Changes to the sources of NumPy and Fortran can be performed directly on the contained source files.

(1) Docker Engine, version 28.0.4 or later.
https://www.docker.com/products/docker-desktop/
(2) Artifact scripts/documentation/sources, version 1.0., contained into: https://doi.org/10.5281/zenodo.20407801
Additionally, the source code for all artifacts is provided. It can be used to modify, re-build and evaluate all experiments on any system supporting all technologies involved: NumPy, Numba,

FORTRAN, and .NET8.0. The following software setup produced the results presented in the paper and may also be used for re-evaluating the artifacts:
(1) OS: Windows 11 Pro 24H2, 64 Bit. https://microsoft.com
(2) Intel® Fortran Compiler (ifx) Version 2025.1.0.
https://www.intel.com/content/www/us/en/developer/tools/oneapi/fortran-compiler-download.html
(3) Visual Studio Build Tools (no Visual Studio IDE required)
https://aka.ms/vs/17/release/vs_BuildTools.exe
(4) Python, version 3.13.2.
https://www.python.org/downloads/
(5) NumPy, version 2.1.3. Installed via pip package manager.
(6) Numba, version 0.61.0. Installed via pip package manager.
(7) .NET SDK 8.0 (including .NET Runtime 8.0)
https://dotnet.microsoft.com/en-us/download/dotnet/8.0
(8) Artifact source code, version 1.0:
Locked release: https://doi.org/10.5281/zenodo.20407801
Development: https://github.com/hokb/decentralized-array-execution-artifacts2026
(9) ILNumerics VP nuget package is fetched and installed automatically when building an artifacts code. It may also be obtained manually, from:
https://www.nuget.org/packages/ILNumerics.Computing

### Datasets / Inputs.

All experiments automatically generate input data of identical shape, type, value, striding and precision. The input arrays *A* and *B* shapes are chosen in a way to perform broadcasting and to include work along small as well as along larger dimensions, reducing a dimension of moderate size (dim #1).
A: [507, 10, 5, 17]<uint32> array
B: [1, 1, 5, 17]<uint32> array
The input data can be changed in the source code of all experiments if all experiments data are manually kept in synch. During evaluation, the artifacts check for identical results from all experiments.

### Installation and Deployment (using docker)

Download and install the Docker Engine (Docker Desktop), using default settings. Download and run the docker image on the evaluating system.

### Installation and Deployment (using sources)

For detailed instructions consult the Readme.md, contained in the artefact package at:
https://doi.org/10.5281/zenodo.20407801
Summary: install the Fortran compiler, Python, pip, .NET SDK. Install NumPy and Numba using pip. Download and unzip the artifacts source code into a local directory [Installdir]. Navigate to [Installdir]/Part1/FORTRAN and compile the FORTRAN sources using the script: Compile_ifx_WIN64.bat. Compile the .NET sources using the dotnet CLI.

### Artifact Execution

For source-based evaluation the program "Part1.exe" is executed. It starts the NumPy, Numba, and (precompiled, see above) FORTRAN experiments. Further, it executes the .NET (CLR) experiments in the same process. Each experiment generates identical input data, measures execution times in chunks of 1000 repetitions each for 10 seconds of executing each experiment and stores all timings and corresponding result arrays into text files. These files are collected by the main .NET "Part1" program afterwards and are merged and transformed to create Figure 4 in the paper.

### Artifact Analysis (incl. Outputs)

Each experiment generates two text files, comprising of the expression result array values and of the measured timings over 10 seconds. The main output is an SVG file containing the merged results corresponding to Figure 4. The files are placed into the output folder, typically, into:
[projectdir]\result\

## *2.2 Computational Artifact* ***A2***

The “Embarrassingly Parallel” Loop artifact *A*2 validates the VP’s ability to automatically optimize simple loop bodies (Figure 5). The speed is compared with manual parallelization techniques.

### Relation To Contributions

Uses the EN to detect parallelism regardless of the program’s structure(C2), automatically schedules tasks (C3), and applies pipelining/fusion (C5).

### Expected Results

The automatic optimizations of the VP (‘ILNumerics vectorized, VP’) should lead to execution speed, higher than or comparable to manual parallelization (‘Parallel For’ in Figure 5).

### Expected Reproduction Time (in Minutes)

Less than 5 min.

### Artifact Execution

Start "Part2a.exe" using the dotnet CLI:
> dotnet run -c Release
It contains the VP optimized and manually optimized version of the experiments loop expression and executes them in sequential order. It then re-generates Figure 5 from the collected timings.

### Artifact Analysis (incl. Outputs)

The timings can be analyzed from the generated output SVG figure "Part2a.svg".

### 2.3 Computational Artifact $\boldsymbol{A3}$

Creates Figure 6: benchmarking a loop body with iteration dependencies.

#### Relation To Contributions

VP respects data dependencies (C2) while parallelizing independent computations (C3) and retaining sequential semantics (C5).

#### Expected Results

The VP should automatically enable partial parallelization of the dependent loop, thus surpassing sequential executions speed. The plot `ILNumerics For loop (VP)' should show significantly better speed compared to the sequential version.

#### Expected Reproduction Time (in Minutes)

Less than 5 min.

#### Artifact Execution

Start "Part2b.exe" using the dotnet CLI:
> dotnet run -c Release
It contains all experiments and executes them in sequential order. It then regenerates Figure 6 from the collected timings.

#### Artifact Analysis (incl. Outputs)

The timings can be analyzed from the generated output SVG figure "Part2b.svg".

### *2.4 Computational Artifact **A4***

$A4$ compares the efficiency of the VP for whole code areas with NumPy and FORTRAN, re-creating Figure 7. $A4$ is found in the folder "Part3" in the artefact package at:
https://doi.org/10.5281/zenodo.20407801
Its handling (setup, execution, output) closely follows → 2.1 Computational Artifact **A1**.

#### Relation To Contributions

This artifact validates the VP holistically across loops and functions (C1/C2), automatic optimizations, scheduling & JIT (C3/C4/C5), confirming hardware-agnostic scalability and maintainability (C6).